\documentclass[aps,twocolumn,showpacs,preprintnumbers,amsmath,amssymb,prl]{revtex4-1}
\usepackage[T1]{fontenc}
\usepackage{graphicx}
\usepackage{esint}
\usepackage{color}
\usepackage[outdir=/home/marcq/papers/proliferation/]{epstopdf}

\makeatletter

\@ifundefined{textcolor}{}
{%
 \definecolor{BLACK}{gray}{0}
 \definecolor{WHITE}{gray}{1}
 \definecolor{RED}{rgb}{1,0,0}
 \definecolor{GREEN}{rgb}{0,1,0}
 \definecolor{BLUE}{rgb}{0,0,1}
 \definecolor{CYAN}{cmyk}{1,0,0,0}
 \definecolor{MAGENTA}{cmyk}{0,1,0,0}
 \definecolor{YELLOW}{cmyk}{0,0,1,0}
}

\newcommand{\showfigures}[1]{{#1}} % pour compiler les figures

\newcommand{\DP}[2]{\ensuremath{\frac{\partial{#1}}{\partial{#2}}}}

\newcommand{\DT}[2]{\ensuremath{\frac{\mathrm{d}{#1}}{\mathrm{d}{#2}}}}

\newcommand{\rate}{\kappa}
\newcommand{\ract}{\kappa_{\textrm{a}}}
\newcommand{\sact}{\sigma_{\textrm{a}}}
\newcommand{\rhod}{\rho_{\mathrm{d}}}
\newcommand{\rhoe}{\rho_{\mathrm{e}}}
\newcommand{\tauh}{\tau_{\mathrm{h}}}
\newcommand{\rhoh}{\rho_{\mathrm{h}}}
\newcommand{\sigp}{\sigma_{\mathrm{p}}}
\newcommand{\etascale}{\tilde{\eta}}
\newcommand{\lscale}{\tilde{l}_{12}}

\makeatother

\usepackage[english]{babel}
\begin{document}

\title{Cell growth, division and death in cohesive tissues:\\ a thermodynamic approach}

\author{Shunsuke Yabunaka}
\email{shunsuke.yabunaka@scphys.kyoto-u.ac.jp}
\affiliation{Fukui Institute for Fundamental Chemistry, Kyoto University, Kyoto, Japan}
\author{Philippe Marcq}
\email{philippe.marcq@curie.fr}
\affiliation{Sorbonne Universit\'es, UPMC Universit\'e Paris 6, 
Institut Curie, CNRS, UMR 168, Laboratoire Physico Chimie Curie, Paris, France
}

\date{May 16, 2017}
%\date{\today}

\begin{abstract}
Cell growth, division and death are defining features of biological tissues
that contribute to morphogenesis. In hydrodynamic descriptions of cohesive 
tissues, their occurrence implies a  non-zero rate of variation of cell density.
We show how linear nonequilibrium thermodynamics allows to express this 
rate as a combination of relevant thermodynamic forces: 
chemical potential, velocity divergence, and activity.
We illustrate the resulting effects of the non-conservation of cell density
on simple examples inspired by recent experiments on cell monolayers, 
considering first the velocity of a spreading front, 
and second an instability leading to mechanical waves.
\end{abstract}

\maketitle

\section{Introduction}
\label{sec:intro}

Biological tissues are assemblies of cells in mutual interaction
\cite{Alberts2008}. 
When cell-cell adhesion is strong and stable, cohesive tissues form 
continuous materials. Smooth mechanical
fields can then be read from experimental data, among which the velocity
\cite{Petitjean2010,Vig2016} and the stress field 
\cite{Ishihara2012,Nier2016}.
Upon suitable coarse-graining over domains comprising several cells
\cite{Bosveld2012}, space-time maps of specific mesoscopic quantifiers 
can also be estimated in tissues, seen either as a cellular material 
(maps of cell area and anisotropy) or as a biomaterial 
(maps of cell division, death, and planar cell polarity).

Tissues differ from inert materials by the occurrence of cell division 
and death \cite{Alberts2008} and by the spontaneous generation of 
internal forces due to the activity of molecular motors and to 
nucleotide-dependent polymerization of cytoskeletal filaments 
\cite{Howard2005}. Since they contribute to morphogenesis
\cite{Wolpert2006,Castanon2011,Suzanne2013,LeGoff2015}, 
cell growth, cell division and cell death
must be included in hydrodynamic descriptions of tissues, in particular when
the time scale considered is larger than a typical cell cycle.

Although the reaction of hydrolysis of adenosine 
triphosphate (ATP) is far from equilibrium
($\Delta \mu \simeq 25 \, \mathrm{k}_{\mathrm{B}} T$ in usual conditions),
linear non-equilibrium thermodynamics \cite{Chaikin2000}
has been shown to describe cytoskeletal mechanics with 
considerable success \cite{Kruse2005,Marchetti2013}.
In the same spirit, we apply linear non-equilibrium thermodynamics to 
a continuous material subject to cell growth, division and death,
for which cell number density is not conserved (Sec.~\ref{sec:thermo}).
For illustrative purposes, we consider a viscoelastic cell monolayer 
in one spatial dimension 
\cite{Vedula2012,Yevick2015,Serra-Picamal2012,Tlili2015PhD,Peyret2016PhD}.
In  Sec.~\ref{sec:front}, we first investigate how cross-coefficients
may modify the velocity of a moving free boundary during tissue expansion,
before discussing in Sec.~\ref{sec:polar} 
an example involving a polar order parameter, and examining the impact 
of cell division on the emergence of mechanical waves. 
Concerning terminology, cell ``growth'' refers to volumetric increase 
(or decrease) at fixed total cell number, and cell ``death'' includes
non-lethal cell delamination from planar tissues.
We do not consider the possible effects of cell division and death 
on tissue rheology, which may become relevant on time scales much larger 
than a typical cell cycle  \cite{Ranft2010}.

\section{Linear nonequilibrium thermodynamics}
\label{sec:thermo}

For simplicity, we consider a finite one-dimensional system of fixed 
size $L$, with spatial coordinate $x \in [0 \; L]$ and time $t$. 
The cell number density field $\rho(x,t)$ obeys a balance equation with 
a source term due to cell growth, division and death,
proportional to $\rho$: 
\begin{equation}
\label{eq:cons:cell}
\partial_{t} \rho + \partial_{x} \left(v\rho\right) = \rate \, \rho \,.
\end{equation}
This equation defines the rate of variation of the cell density $\rate$, 
which we shall determine within the framework of linear nonequilibrium 
thermodynamics. We denote the tissue velocity and stress fields $v(x,t)$ and
$\sigma(x,t)$, respectively.
In the presence of an external force field $f_{\mathrm{ext}}$,
the conservation of linear momentum reduces to the force balance equation:
\begin{equation}
\label{eq:cons:momentum}
\partial_x \sigma = - f_{\mathrm{ext}} \,,
\end{equation}
since inertia is negligible at the length and velocity scales
characteristic of tissue mechanics.
We consider isothermal transformations at a constant, uniform 
temperature $T$. Given $f(\rho)$ the free energy density per unit length, 
we deduce the chemical potential $\mu=  \left( \DP{f}{\rho} \right)_T$,
and the pressure field $\pi =  -f + \mu \rho$.

Eqs.~(\ref{eq:cons:cell}-\ref{eq:cons:momentum}) are supplemented by
balance equations for the energy density $u$ and entropy density $s$:
\begin{eqnarray}
  \label{eq:bal:energy}
  \DT{u}{t} &=& - u \partial_x v -\partial_x j^u + f_{\mathrm{ext}} v \,,\\
  \label{eq:bal:entropy}
  \DT{s}{t} &=& - s \partial_x v - \partial_x j^s + \Sigma \,,
\end{eqnarray}
including density, energy and entropy currents 
$j^{\rho}$,  $j^u$ and $j^s$ and the
entropy production rate $\Sigma$. In 1D the total derivative is
$\DT{}{t} = \DP{}{t} + v \DP{}{x}$. 
From the thermodynamic equality  
$\mathrm{d}u = T \mathrm{d}s + \mu \mathrm{d}\rho$, and identifying the 
pressure $\pi = - u + \mu \rho +T s$, we obtain
\begin{eqnarray}
  T \, \Sigma % &=&  T \left( \DT{s}{t} + s \partial_x v + \partial_x j^s \right) \\
&=& \DT{u}{t} - \mu \DT{\rho}{t} + T s \partial_x v + T \partial_x j^s \,,\\
&=& \partial_x J - \mu \rate \rho
+ \partial_x v \left( \sigma + \pi \right)
% \left( \sigma - u + \mu \rho +T s \right)
- j^{\rho} \, \partial_x \mu \,,
\end{eqnarray}
using (\ref{eq:cons:cell}-\ref{eq:bal:entropy}) and  
integrations by parts. The current $J = -j^u + T j^s 
+ \mu j^{\rho} - v \sigma$
contributes through a boundary term, and
may thus be ignored in bulk. 
Among possible fluxes and forces, only 
$j^{\rho}$ and $\partial_x \mu$ change sign under the transformation $x \to -x$.
At linear order, this forbids possible cross-couplings between
$j^{\rho}$, $\partial_x \mu$ and other forces and fluxes. 
The diagonal term leads to Fickian diffusion \cite{Kruse2005},
$j^{\rho} = - D \,\partial_x \rho$, irrelevant in the case of
cohesive tissues. We therefore neglect the flux-force pair
$( j^{\rho} , - \partial_x \mu)$ from now on.

The hydrolysis of ATP is schematically 
represented as ATP $\to$ ADP $+ \mathrm{P}_{\mathrm{i}}$. It 
proceeds at rate $r$, for a variation of chemical potential 
$\Delta \mu = \mu_{\mathrm{ATP}} - \mu_{\mathrm{ADP}} - \mu_{\mathrm{P}_{\mathrm{i}}}$,
assumed to be constant. 
Taking into account ATP hydrolysis, we 
obtain the dissipation rate as the sum of thermodynamic flux-force products:
\begin{equation}
  \label{eq:entprod:scalar}
  T \Sigma = \rate \left(-\rho\mu\right) 
+ \left(\sigma+\pi\right)\partial_{x}v + r\Delta\mu  \,.
\end{equation}
Despite its relevance for models of a dynamic, polymerizing and 
depolymerizing cytoskeleton, 
the term $\rate \left(-\rho\mu\right)$ has not been studied explicitly in 
models of active matter \cite{Kruse2005,Doostmohammadi2015}.
Since $\left(-\rho\mu\right)$  can be computed from the free energy density,
we treat it as a thermodynamic force and define the following flux-force pairs:
\begin{equation*}
\begin{array}{ccc}
\mathrm{Flux} & \leftrightarrow & \mathrm{Force}\\
\rate & \leftrightarrow & -\rho\mu\\
\left(\sigma+\pi\right) & \leftrightarrow & \partial_{x}v\\
r & \leftrightarrow & \Delta\mu
\end{array}
\end{equation*}
As discussed in Appendix~\ref{sec:app:scalar}, choice of fluxes and forces 
has some arbitrariness in linear non-equilibrium thermodynamics, but this does 
not lead to essential differences in the resulting hydrodynamic equations.

To linear order, the constitutive equations read
\begin{eqnarray}
\label{eq:const:scalar:kd}
\rate&=& l_{11} \, \left(-\rho\mu\right)+ l_{12} \, \partial_{x}v+\ract\\
  \label{eq:const:scalar:stress}
\sigma+\pi &=& - l_{12} \, \left(-\rho\mu\right) 
+ \eta \, \partial_{x}v +\sact \,,
\end{eqnarray}
where the diagonal coefficients $l_{11}$ and $l_{22}$ %and $l_{33}$ 
are non-negative, and Onsager relations have been applied.
Since $r$ is not easily measurable, we ignore the analogous equation 
relating $r$ to the same forces.
We recognize $\eta = l_{22}$ as the tissue viscosity, and 
$\sact = l_{23} \, \Delta \mu$ as the active stress \cite{Kruse2005}.
The subscript $\mathrm{a}$ indicates that a parameter is active. 
We define an active rate $\ract = l_{13} \, \Delta \mu$,
which may be understood as a ``swelling'' rate \cite{Tlili2015},
negative (respectively positive) when cell volume increases  
(respectively decreases). In the presence of cell growth, we define
the homeostatic density $\rhoh$ as the density at which cell growth, 
division and death balance each other in the absence of flow, \emph{i.e.}
$\kappa(\rhoh) = 0$ with $v = \partial_x v = 0$.
Just as the active stress  shifts the reference density
at which the stress vanishes in the absence of viscous dissipation,
the active rate shifts the homeostatic density (see 
Eqs.~(\ref{eq:const:front:kd}-\ref{eq:const:front:sigma}) for an example).
Another approach \cite{Tlili2015} posits $\rate$ as the sum of three rates,
of cell growth, cell division, and cell death respectively. Since each 
process may be regulated by cell density 
and/or velocity divergence, and requires ATP hydrolysis for its completion, 
we expect the three rates to be functions of $\rho$, $\partial_x v$ and 
$\Delta \mu$. Only their sum $\rate$ can be specified unambiguously
by thermodynamics, Eq.~\eqref{eq:const:scalar:kd}.  

The division rate has been observed experimentally to decrease with 
$\rho$ \cite{Folkman1978,Puliafito2012,Streichan2014}.
When divisions dominate $\kappa$, the positivity of $l_{11}$ implies
that the chemical potential is negative and increases monotonically 
with $\rho$ (see examples below).
For large values of the density (``overcrowding''), the chemical
potential may become positive, whereby $\kappa$ becomes negative,
indicating that cell delaminations dominate 
\cite{Eisenhoffer2012,Marinari2012}.
Through $-\rho \mu(\rho)$, $\kappa$ depends implicitly on 
the pressure $\pi(\rho)$, 
as proposed and investigated in the non-linear regime in \cite{Shraiman2005}.
We expect $\kappa$ to be a decreasing function of pressure, 
as found experimentally to be caused either by an enhanced
apoptotic rate  \cite{Helmlinger1997} or by a reduced division
rate  \cite{Delarue2014}, or by both \cite{Cheng2009}. 
The rate of cell division also correlates with tissue contractility,
while inhibitors of contractility alter spatial patterns of proliferation
\cite{Nelson2005}. 

\begin{figure}[!t]
\showfigures{
\includegraphics[scale=0.15]{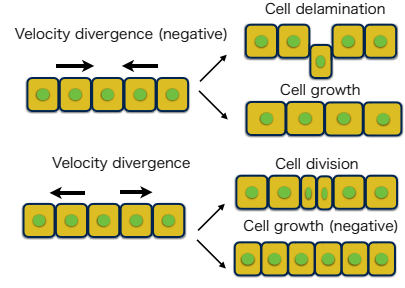}
}
\vspace*{-0.5cm}
\caption{
\textbf{Schematic representation}  of the effect of the  dimensionless 
cross-coefficient $l_{12}$ that couples cell growth, cell division and 
cell death to the velocity divergence ($l_{12}>0$).
}
\label{fig:illustration}
\end{figure}

Next, we explain the significance of the newly introduced  dimensionless 
cross-coefficient $l_{12}$. Firstly, it couples
the divergence of the tissue velocity to $\kappa$ 
as illustrated in Fig.~\ref{fig:illustration}. 
Observations of a positive correlation between 
a negative velocity divergence (``tissue convergence'') and cell
delaminations (negative $\kappa$) \cite{Levayer2016} suggest that 
$l_{12} \ge 0$. 
Neglecting the influence of 
growth, divisions, and cell density variation, we roughly estimate  
$l_{12}\approx 10^{1}$ from the measurements
of the cell delamination rate and tissue convergence
in the fruitfly pupal midline \cite{Levayer2016}.
Secondly, substituting the expression of $-\rho\mu$ obtained from 
\eqref{eq:const:scalar:kd} into \eqref{eq:const:scalar:stress} 
allows to rewrite  the stress field as:
\begin{equation}
  \label{eq:const:stress}
\sigma = -\pi - \frac{l_{12}}{l_{11}} \rate 
+ \left( \eta + \frac{l_{12}^2}{l_{11}} \right) \, \partial_{x}v 
+\sact + \frac{l_{12}}{l_{11}} \ract \,,
\end{equation}
implying that, through the cross-coefficient $l_{12}$, cell growth, 
division and death may modify the tissue mechanical behaviour, 
by changing its pressure, its viscosity and its active stress. 
In particular, we predict that $l_{12}$ increases the effective viscosity
$\eta_{\mathrm{eff}} = \eta + \frac{l_{12}^2}{l_{11}}$ in general,
and the effective pressure 
$\pi_{\mathrm{eff}} = \pi + \frac{l_{12}}{l_{11}} \rate$ when $\kappa \ge 0$. 
A similar (shear) stress contribution due to cell divisions
in 2D has been introduced phenomenologically to explain anisotropic growth 
in the fruitfly wing disk \cite{Bittig2009}.

\section{First example: front velocity of an expanding cell monolayer}
\label{sec:front}

To illustrate the thermodynamic approach by a first concrete, yet 
simple example, let us consider the spreading of a cell monolayer in a 
quasi one-dimensional geometry, either within a channel \cite{Vedula2012}, 
or along a linear fiber \cite{Yevick2015}, and denote $L(t)$ its spatial 
extension at time $t$. Since it involves a free, moving boundary, this
calculation is relevant to wound healing assays performed over
long enough durations \cite{Murray2002}. 
The monolayer is compressible since the cell density
decreases monotonically along $x$, and goes to zero at
the free boundary, $x = L(t)$. Following \cite{Recho2016}, 
we introduce the free energy density,
\begin{equation}
\label{eq:front:f}
f\left(\rho\right)=  E \left(\log\left(\frac{\rhoe}{\rho}\right)
+ \frac{\rho}{\rhod} - 1\right)
\end{equation}
associated with two distinct reference cell densities $\rhoe$ and $\rhod$.
We deduce the chemical potential 
$\mu(\rho) =  \frac{E}{\rho} \left(\frac{\rho}{\rhod}  - 1\right)$,
the pressure field $\pi(\rho) = E \, \log\left(\frac{\rho}{\rhoe}\right)$,
and interpret in 1D the coefficient $E$ as an elastic modulus and
$\rhoe$ as a reference elastic density.
The product $l_{11}E$ has the dimension of inverse time.
Since $E\approx10^{3} \,\mathrm{Pa}$ \cite{Harris2012}
and  $\tauh^0\approx10^{4} \, \mathrm{s}$ \cite{Puliafito2012}
we expect an order of magnitude for 
$l_{11}\approx10^{-7}\, \mathrm{Pa^{-1}s^{-1}}$.

The external force $f_{\mathrm{ext}} = - \xi v$
is dissipative, with a positive friction coefficient $\xi$.
An additional ingredient is the active boundary stress 
$\sigp = \sigma\left(x=L(t), t\right)$,
generated by the lamellipodial activity of leading cells,
and assumed to be constant for simplicity.
In the absence of cross-couplings ($l_{12} = \ract = \sact = 0$), 
the ``bare'' homeostatic stress 
$\sigma_{h}^0 = - E \log\left(\frac{\rhod}{\rhoe}\right)$
is observed in bulk, where the tissue is under tension when $\rhod < \rhoe$.
Altogether, the free front may be pushed or pulled 
depending on the sign of the dimensionless control parameter
\begin{equation}
\label{eq:alpha:0}
\alpha_0 =  \frac{\sigp-\sigma_{h}^0}{E} =
\frac{\sigp}{E} + \log\left(\frac{\rhod}{\rhoe}\right) \,,
\end{equation}
leading to front motion at constant velocity $V$ \cite{Recho2016}.

From (\ref{eq:const:scalar:kd}-\ref{eq:const:scalar:stress}-\ref{eq:front:f}),
the constitutive equations read:
\begin{eqnarray}
  \label{eq:const:front:kd}
\rate  &=&  l_{11}E \, \left(1-\frac{\rho}{\rhod}\right) 
+l_{12} \, \partial_{x}v + \ract \\
  \label{eq:const:front:sigma}
\sigma &=&  - E\log\left(\frac{\rho}{\rhoe}\right)
+ l_{12}E \, \left( \frac{\rho}{\rhod} -1 \right) 
+ \eta\partial_{x}v +  \sact 
\end{eqnarray}
The product $l_{11}E$ has the dimension of inverse time.
We define the dimensionless active rate
$\tilde{\kappa}_{\mathrm{a}}={\ract}/{l_{11} E}$.
A non-zero active rate $\ract$ shifts the homeostatic density
 $\rhoh$ from its  ``bare'' value $\rhoh^0 = \rhod$ to  
$\rhoh = \rhod \, \left( 1 + \tilde{\kappa}_{\mathrm{a}} \right)$,
and the associated characteristic time $\tauh$ from 
$\tauh^0 = \left( l_{11} E \right)^{-1}$ to $\tauh = \left( l_{11} E \right)^{-1} 
\left( 1 + \tilde{\kappa}_{\mathrm{a}} \right)^{-1}$.
In the general case, the control parameter reads (see below)
\begin{equation}
\label{eq:alpha:1}
\alpha= \frac{\sigp - \sact}{E} + \log\left(\frac{\rhod}{\rhoe}
\left( 1 + \tilde{\kappa}_{\mathrm{a}}  \right)\right) 
- \tilde{\kappa}_{\mathrm{a}} l_{12} \,.
\end{equation}

Although a detailed study of the influence of the parameters $l_{12}$
and $\ract$ on front propagation at arbitrary driving is beyond the scope 
of this work, we extend a perturbative calculation of the front velocity 
$V$ done in \cite{Recho2016} for $l_{12}= \ract = 0$.
For convenience, we introduce the following dimensionless quantities: 
$\hat{\rho}=\frac{\rho}{\rhoh}$, $\hat{t}=\frac{t}{\tauh}$
$\hat{x}= x \sqrt{\frac{\xi}{E \tauh}}$ and 
$\hat{\sigma} = (\sigma -\sigp)/E$.
The continuity and force balance equation now read
\begin{eqnarray}
\label{eq:front:rho:dimless}
&& \partial_{t}\hat{\rho}+\partial_{\hat{x}}\left(\hat{v}\hat{\rho}\right) =
\hat{\rho}\left(1-\hat{\rho}\right)  + 
l_{12}\,\partial_{\hat{x}}\hat{v} \\
%\label{eq:front:force:dimless}
&& \partial_{\hat{x}}\hat{\sigma} = \hat{v} 
% \\
% && 
\label{eq:front:sigma:dimless}
\end{eqnarray}
with $\hat{\sigma} = - \log(\hat{\rho}) + \lscale \left( \hat{\rho} - 1\right)
+ \etascale \, \partial_{\hat{x}}\hat{v} - \alpha$, and
the definitions $\etascale = \frac{\eta}{E \tauh}$, 
$\lscale = l_{12} \frac{\rhoh}{\rhod}$, and where $\alpha$ is given by 
\eqref{eq:alpha:1}.
Hereafter we shall omit the hats on dimensionless quantities
(but not the tildes on dimensionless parameters).
The boundary conditions for the scaled stress field are given by
$\sigma\left(x=L\left(t\right),t\right)=0$, 
$\partial_{x}\sigma\left(x=L(t),t\right)=\dot{L}(t)$,
$\partial_{x}\sigma\left(x=0,t\right)=0$.

We shall solve for the front velocity $V$ in steady front propagation
for small $|\alpha|$. We denote the front position by
$z=0$, where $z$ is defined as  $z=x-Vt$.
The conservation equations are 
\begin{eqnarray}
%\label{eq:pert:rho:0}
\nonumber
-V\rho'+\left(\rho \sigma'\right)' &=& 
\rho\left(1-\rho\right)+l_{12} \,  \sigma''\\
%\label{eq:pert:sig:0}
\nonumber
\etascale \, \sigma''-\sigma &=&
\log\rho+\lscale \left(1-\rho\right) + \alpha\,,
\end{eqnarray}
where a $'$ denotes the derivative with respect to $z$, and
the boundary conditions become
$\sigma\left(z = 0\right)=0$, 
$\sigma'\left(z = 0\right)=V$,
$\lim_{z \to - \infty}\sigma'\left(z\right)=0$.

We expand all variables and fields around the steady state obtained when 
$\alpha=0$: $\alpha=0+\epsilon\alpha_1$, $V=0+\epsilon V_1$,
$\sigma(z)=0+\epsilon \sigma_1(z)$, $\rho(z)=1+\epsilon\rho_1(z)$.
At order $\epsilon^1$, we find:
\begin{eqnarray}
\nonumber
%\label{eq:pert:rho:1}
\sigma_1'' &=&-\rho_1 +l_{12} \, \sigma_1'' \\
\nonumber
\etascale \sigma_1''- \sigma_1 &=&\rho_1  (1 - \lscale ) +\alpha_1 
\end{eqnarray}
and deduce the following differential equation for $\sigma_1$: 
\begin{equation}
\label{eq:pert:sig:1}
\left(\etascale + (1-l_{12}) (1-\lscale ) \right) \, \sigma_1'' -\sigma_1 = \alpha_1\,.
\end{equation}
As shown below, the quantity $\etascale + (1-l_{12}) (1-\lscale)$ must 
be positive for linear stability of the uniform bulk state 
$\rho=\rhoh, V=0$. 
A perturbation of small amplitude with wave number $q$ reads 
\begin{equation}
\nonumber
\left(\rho(x,t),\sigma(x,t)\right) =
\left(1,0\right)+\left(\delta\rho,\delta \sigma\right)\, e^{st-iqx}
\end{equation}
with a growth rate $s$.  Using 
Eqs.~(\ref{eq:front:rho:dimless}-\ref{eq:front:sigma:dimless}), 
we find at linear order, 
\begin{eqnarray}
%\label{eq:pert:rho:01}
\nonumber
s\delta \rho-q^{2}\delta\sigma &=& 
-\delta \rho-q^{2}l_{12} \,  \delta\sigma \,,\\
%\label{eq:pert:sig:01}
\nonumber
-q^{2} \etascale \, \delta \sigma-\delta \sigma &=&
(1-\lscale) \delta\rho \,,
\end{eqnarray}
and determine the growth rate as
\begin{equation}
\nonumber
s(q)=-\frac{\tilde{\eta}+(1-l_{12}) (1-\lscale)+1/q^{2}}{\tilde{\eta}+1/q^{2}} \,.
\end{equation}
Linear stability ($s(q) < 0, \forall q$) indeed requires 
the positivity of $\tilde{\eta}+(1-l_{12}) (1-\lscale) $.

Solving \eqref{eq:pert:sig:1}, we obtain the expression 
of the stress profile
\begin{equation}
\label{eq:app:profile:stress}
\sigma_1(z)= \alpha_1 \left(
e^{z / \sqrt{\tilde{\eta} + (1-l_{12}) (1-\lscale )} }-1
\right) \,,
\end{equation}
from which we deduce the velocity and cell density profiles:
\begin{eqnarray}
\label{eq:app:profile:velocity}
v_1(z) &=&  \frac{\alpha_1  }{\sqrt{\tilde{\eta} + (1-l_{12}) (1-\lscale )}} e^{z / \sqrt{\tilde{\eta} + (1-l_{12}) (1-\lscale )} }
\qquad\\
\label{eq:app:profile:density}
\rho_1(z) &=&  \frac{\alpha_1  \,(l_{12}-1)}{\tilde{\eta} + (1-l_{12}) (1-\lscale )} e^{z / \sqrt{\tilde{\eta} + (1-l_{12}) (1-\lscale )} }
\,.
\end{eqnarray}

The front velocity 
is calculated from the boundary condition  $V_1 = v_1\left(z = 0\right)$. 
We find that in the limit of small driving $|\alpha| \ll 1$ and when 
$l_{12} \neq 0$, 
$\ract \neq 0$, the dimensionless front velocity reads
\begin{equation}
\label{eq:front:V}
\mathcal{V} 
= \frac{V}{\sqrt{\frac{E}{\xi\, \tauh^0}}}
=  \alpha \, \,
       \sqrt{ \frac{1 + \tilde{\kappa}_{\mathrm{a}}}{ 
           \etascale +\left(1-l_{12}\right) 
    \left(1- l_{12} \left( 1 + \tilde{\kappa}_{\mathrm{a}} \right)  \right) 
          } }  \, ,
\end{equation}
as a function of $\alpha$, $l_{12}$, $ \tilde{\kappa}_{\mathrm{a}}$, and
$\etascale  = \frac{\eta}{E \tauh}$ the dimensionless viscosity.
Since a finite homeostatic density
requires $1 + \tilde{\kappa}_{\mathrm{a}} > 0$ 
(Eq.~\eqref{eq:const:front:kd}),
the argument of the square root in \eqref{eq:front:V} is positive. 

\begin{figure}[!t]
\showfigures{
\includegraphics[scale=0.2]{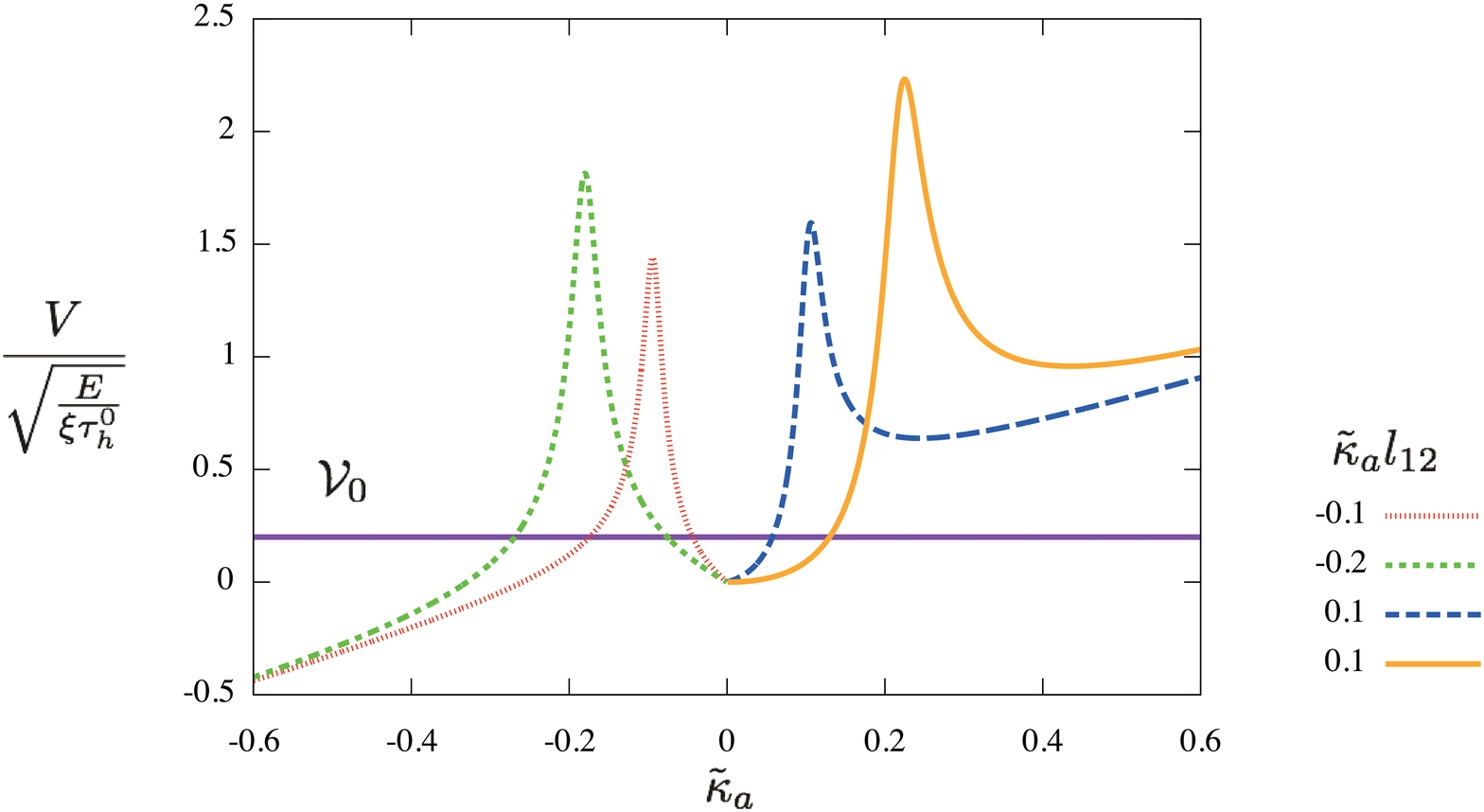}
}
\caption{\textbf{Dimensionless front velocity}
$\mathcal{V} = V/\sqrt{\frac{E}{\xi\, \tauh^0}}$ as a function of 
$\tilde{\kappa}_{\mathrm{a}}$ at fixed 
$\tilde{\kappa}_{\mathrm{a}} l_{12} = -0.2, -0.1, 0.1, 0.2$,
assuming $l_{12} \ge 0$,
with $\alpha_0 = 0.2$ and $\etascale = 10^{-2}$.
The solid line indicates the reference velocity
$\mathcal{V}_0 %= \alpha_0/\sqrt{1 + \etascale} \simeq \alpha_0 
= 0.2$ when $l_{12} = \ract = 0$.
Using $E \approx 10^3 \,{\mathrm Pa}$ \cite{Harris2012},
$\xi \approx 10^{16} \, \mathrm{Pa}\,\mathrm{m}^{-2}$ s 
\cite{Cochet-Escartin2014} and
$\tauh^0 \approx 10^4$ s \cite{Puliafito2012}, the velocity scale is 
$\sqrt{\frac{E}{\xi\, \tauh^0}} \approx 10 \, \mu\mathrm{m} \, \mathrm{h}^{-1}$ \cite{Recho2016}.
See also Fig.~\ref{fig:V2} for the case $l_{12} < 0$.
}
\label{fig:V}
\end{figure}

When $\ract = 0$, the driving $\alpha$ does not depend on $l_{12}$,
which can adopt arbitrary large values while $|\alpha| \ll 1$.
Since $\etascale \approx 10^{-2}$ \cite{Recho2016}, the front velocity 
is reduced by a factor close to $|l_{12}|$ when $|l_{12}| \gg 1$ and 
$\ract = 0$. When $\ract \neq 0$, the driving $\alpha$ may remain small
provided that $\ract l_{12}$ is also small. 
Fig.~\ref{fig:V} shows how $\mathcal{V}$
depends on the active rate $\ract$ at fixed, small $\ract l_{12}$, when
$l_{12} \ge 0$ and $\alpha_0 = 0.2$ (reference velocity 
$\mathcal{V}_0 \simeq \alpha_0 = 0.2$
when $l_{12} = \ract = 0$). A large enough, positive $\ract$ increases 
$\mathcal{V}$ above $\mathcal{V}_0$, with a maximal value 
$\mathcal{V}_{\mathrm{max}} \gg \mathcal{V}_0$ reached close to
$l_{12} = 1$. A large, positive maximal velocity is also obtained for negative
$\ract$ close to $l_{12} = 1$. Remarkably, a negative $\ract$ may 
change the sign of the velocity as $\alpha$ becomes negative. 
We  conclude that the sign and numerical value of the front velocity 
are sensitive to the cross-coefficients $l_{12}$ and $\ract$, while
$l_{11}$ determines the velocity scale. 

The above expression of the front velocity \eqref{eq:front:V}, 
together with the profiles of stress, velocity and cell density, 
Eqs.~(\ref{eq:app:profile:stress}-\ref{eq:app:profile:density}) 
can be tested experimentally.
Comparison with spreading assays where either cell division, cell apoptosis, 
and/or contractility are inhibited may lead to 
quantitative estimates of $l_{12}$ and $\tilde{\kappa}_{\mathrm{a}}$.

\section{Second example: mechanical waves in a polar tissue}
\label{sec:polar}

As a second example, we ask how pattern formation in a polar tissue
is modified by cell growth, division and death, or more precisely
how the location of bifurcation thresholds leading to wave patterns
depends on $l_{11}$ and $l_{12}$.
Active gel models are prone to instabilities driven by their 
active coefficients \cite{Kruse2005,Marchetti2013}.
Experimentally, propagating mechanical waves have been observed close to the
moving boundary of expanding epithelial monolayers 
\cite{Serra-Picamal2012,Tlili2015PhD},
as well as in the bulk of confined systems \cite{Peyret2016PhD},
over time scales similar to or larger than the typical cell cycle. 
Whereas other models of an instability leading to  mechanical 
waves  consider an incompressible material
\cite{Blanch2014,Banerjee2015,Notbohm2016}, 
we note that epithelial cell monolayers are 
compressible in 1D or 2D, with large fluctuations of cell sizes 
\cite{Serra-Picamal2012,Zehnder2015a,Zehnder2015b}. This observation
justifies Eq.~\eqref{eq:cons:cell}.

In motile cells, cell polarity arises from the distinct morphology of 
front and rear, from the inhomogeneous profiles of signaling molecules 
such as Rho and Rac, or from the respective positions of the cell 
centrosome and nucleus \cite{Mao2016}. 
Coarse-graining cell polarity at the tissue scale,
we take into account a smooth polarity field $p(x,t)$  to
describe the collective motion of a cohesive cell assembly. 
The constitutive equations of a polar material 
involve an additional flux-force pair $\dot{p} \leftrightarrow  h$
\cite{Kruse2005}, where $h$ is the field conjugate to  $p$  and
$\dot{p} = \partial_t + v \partial_x p$, see Eq.~\eqref{eq:R:polar}. 
Given the polar invariance of the tissue
under $p \to -p$, $x \to -x$, 
Eqs.~(\ref{eq:const:scalar:kd}-\ref{eq:const:scalar:stress}) 
generalize to 
\begin{eqnarray}
\label{eq:const:polar:kd}
\rate&=& l_{11} \, \left(-\rho\mu\right)+ l_{12} \, \partial_{x}v+\ract
+\gamma_{\mathrm{a}}\partial_{x}p\\
  \label{eq:const:polar:stress}
\sigma+\pi &=& - l_{12} \, \left(-\rho\mu\right) 
+ \eta \, \partial_{x}v +\sact +\beta_{\mathrm{a}}\partial_{x}p\\
  \label{eq:const:polar:p}
\dot{p}&=& \Gamma_{\mathrm{p}}h 
\end{eqnarray}
where $\Gamma_{\mathrm{p}} \ge 0$. The active parameters
$\beta_{\mathrm{a}}$ and $\gamma_{\mathrm{a}}$
couple thermodynamic fluxes to the polarity divergence.

To study quantitatively the mechanical waves 
observed in expanding tissues \cite{Serra-Picamal2012,Tlili2015PhD}, 
one would need to combine both examples, \emph{e.g.} associating
the stress boundary condition at $x = L(t)$ to this  analysis. 
Here, we focus on the question of how the emergence of waves is influenced 
by cell growth, division and death in bulk, and consider
a system of fixed length $L$ with periodic boundary conditions,
as may be realised in an annular geometry \cite{Nier2016}. 
A minimal expression for the free energy density reads
\begin{equation}
\label{def:polar:f}
f= \psi_{\rho}\left(\rho\right) +
\psi_{\mathrm{p}}\left(p\right)  
+w \rho \, \partial_{x}p
+\frac{\nu_4}{2}\left(\partial_{x}^{2}p\right)^{2}
\end{equation}
including a quadratic function of the density
\begin{equation}
\label{def:polar:psirho}
\psi_{\rho}\left(\rho\right) = 
\frac{1}{2K} \left(\frac{\rho-\rhoe}{\rhoe}\right)^{2}
\end{equation}
that sets the reference elastic density $\rhoe$,
with a compressibility coefficient $K$; and a polarity-dependent term
\begin{equation}
\label{def:polar:psip}
\psi_{\mathrm{p}}\left(p\right)=-\frac{a_{2}}{2}p^{2}+\frac{a_{4}}{4}p^{4}
\end{equation}
with $a_{2}, a_{4} \ge 0$, that sets the reference polarity
$p_0 = \sqrt{a_2/a_4}$. As allowed by symmetry,
the term $w \rho \, \partial_{x}p$ in \eqref{def:polar:f} couples
cell density and polarity divergence with a coefficient $w$ of unspecified 
sign \cite{Marcq2014}.
The last term in \eqref{def:polar:f} suppresses the 
instability at large wave numbers ($\nu_4 \ge 0$) \cite{Chaikin2000}. 
Following  \cite{Tlili2015PhD,Blanch2014,Banerjee2015,Notbohm2016},
we include an active motility term in the external force 
$f_{\mathrm{ext}} = - \xi v + f_{\mathrm{a}} p$ with a positive 
coefficient $f_{\mathrm{a}}$. 

In the homogeneous state $p(x,t) = p_0$, 
$v(x,t) = v_{0}=f_{\mathrm{a}} p_{0}/\xi$,
the density $\rho_{0}$ is determined by solving $\rate(\rho_0) = 0$ 
or $\rho_0\,\mu(\rho_0) = \ract/l_{11}$. For simplicity, we set 
$\ract = 0$, so that $\rho_0 = \rhoe$,
and consider small perturbations  around
$(\rhoe, p_{0}, v_0)$, see Eq.~\eqref{eq:linear:polar:pert}
in Appendix~\ref{sec:app:stab}.
The growth rate of the instability is determined numerically from 
Eqs.~(\ref{eq:linear:polar:l1}-\ref{eq:linear:polar:l3}).
We find that the primary instability is a Hopf 
bifurcation, leading to traveling waves. %, and that

\begin{figure}[!t]
\showfigures{
\includegraphics[scale=0.5]{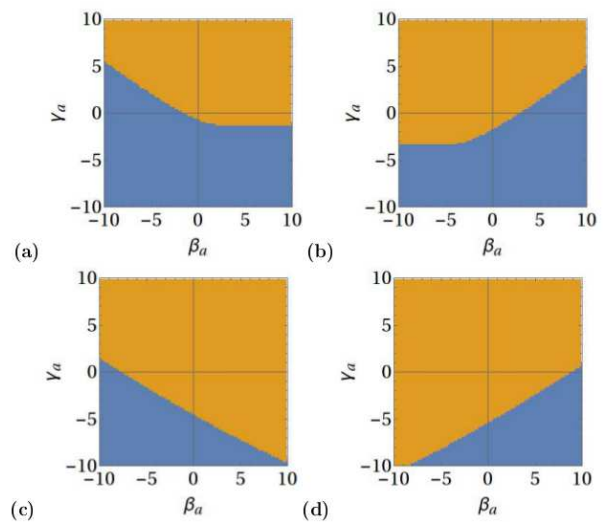}
}
\caption{
\textbf{Stability diagram} in the plane $(\beta_{\mathrm{a}}, \gamma_{\mathrm{a}})$ 
of active control parameters for 
(a) $l_{11} = l_{12} = 0$; 
(b) $l_{11} = 0$, $l_{12} = 2$; 
(c) $l_{11} = 2$, $l_{12} = 0$; 
(d) $l_{11} = l_{12} = 2$. 
Orange (respectively blue) corresponds to a stable 
(respectively unstable) uniform state. Parameter values are: 
$\ract = 0$, $\eta = \sact = \Gamma_p =  w = \nu_4 = K = \rhoe = a_2 =  
\xi = f_{\mathrm{a}} = 1$ and $p_0=0.5$. 
See also Fig.~\ref{fig:bif:2} for the case $l_{12} = -2$. 
}
\label{fig:bif}
\end{figure}

With respect to the active control parameters $\beta_{\mathrm{a}}$ and
$\gamma_{\mathrm{a}}$, bifurcation thresholds are sensitive to 
the coefficients $l_{11}$ and $l_{12}$ (see Fig.~\ref{fig:bif}).
In particular, we find that the instability is suppressed due 
to $l_{11}$. In the vanishing wavenumber limit, 
density perturbations obey $\left(s+ \frac{l_{11}}{K} \right)\delta\rho = 0$
(see Eq.~(\ref{eq:linear:polar:l1})), and decouple from pressure and 
velocity perturbations. One solution for the growth rate is 
$s = - \frac{l_{11}}{K} < 0$, suggesting that the source term 
in \eqref{eq:cons:cell} stabilizes the uniform state through 
the coefficient $l_{11}$. 
Experimentally, pharmacological inhibition of cell division enhances
waves in expanding monolayers  \cite{Tlili2015PhD}, in accord with 
our model. Since division and death are treated
through the same field $\rate$, this further suggests that inhibiting 
cell death would also enhance waves.

When $l_{11} = l_{12} = 0$, Fig.~\ref{fig:bif}a gives the bifurcation line 
in the $(\beta_{\mathrm{a}}, \gamma_{\mathrm{a}})$ plane. Setting  
$l_{11} = 0$, $l_{12} = 2$ (Fig.~\ref{fig:bif}b), the instability now occurs 
\emph{above} a threshold value of $\beta_{\mathrm{a}}$ at fixed 
$\gamma_{\mathrm{a}}$, instead of \emph{below} a threshold when 
$l_{11} = l_{12} = 0$ 
(compare also Figs.~\ref{fig:bif}c and \ref{fig:bif}d). 
An analytical calculation performed in  the simpler case 
$\eta = l_{11} = f_{\mathrm{a}} = \gamma_{\mathrm{a}}=0$
(Appendix~\ref{sec:app:stab:alpha0})
predicts that the bifurcation diagram depends 
on the sign of $w (1 - l_{12})$, and thus on whether $l_{12}>1$ or $l_{12}<1$
(see Eq.~\eqref{eq:linstab:simple:cond}).
In the general case, we observed numerically that smaller 
(respectively larger)
$\beta_{\mathrm{a}}$ is favorable for the instability when $w(1 - l_{12}) >0$
(respectively $w(1 - l_{12}) <0$).

\begin{figure}[!t]
\showfigures{
\includegraphics[scale=0.45]{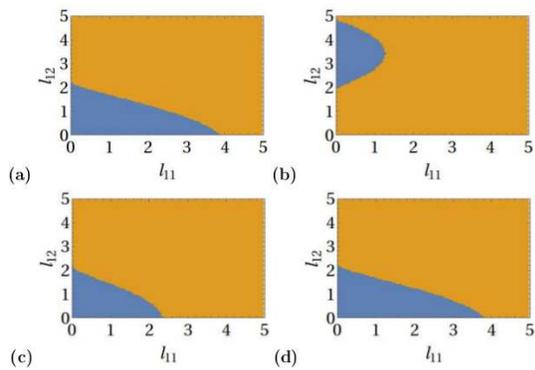}
}
\caption{\textbf{Stability diagrams} in the plane  $(l_{11} , l_{12})$: 
(a) $\beta_{\mathrm{a}} = \gamma_{\mathrm{a}} = -5$;
(b) $\beta_{\mathrm{a}} =10$, $\gamma_{\mathrm{a}} = 5$.
The color code and other parameter values are as in Fig.~\ref{fig:bif}.
See also Fig.~\ref{fig:bif:3} for the case $l_{12} < 0$.
Parameters in (c,d) are the same as in (a), with additional terms
(c) $\alpha_{\mathrm{a}} =0, \nu_2 = 1$;
(d) $\alpha_{\mathrm{a}}=1,\nu_2=0$.
}
\label{fig:bif:l}
\end{figure}

\begin{figure}[!t]
\showfigures{
\includegraphics[scale=0.25]{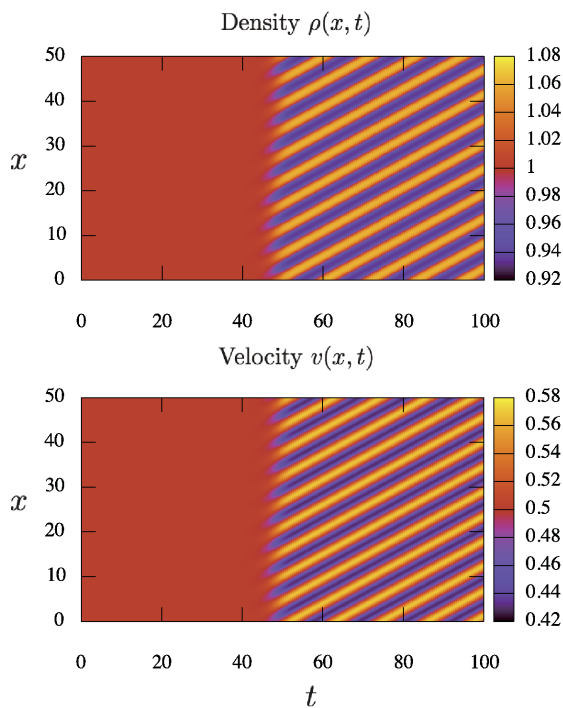}
}
\caption{
\textbf{Numerical simulation} 
Density and velocity fields obtained by numerical resolution 
of  Eqs.~(\ref{eq:cons:cell}-\ref{eq:cons:momentum},
\ref{eq:const:polar:kd}-\ref{eq:const:polar:p}).
Initial parameters $(l_{11},l_{12})=(1,3)$ are switched to 
$(l_{11},l_{12})=(1,1)$ when $t = 10$ to induce an instability of the 
uniform state to a traveling wave. 
Other parameters are the same as in Fig.~\ref{fig:bif:l}a.
}
\label{fig:bif:num}
\end{figure}

\begin{figure}[!t]
\showfigures{
\includegraphics[scale=0.5]{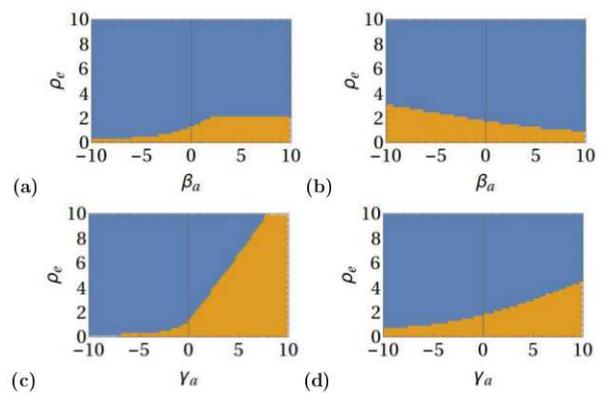}
}
\caption{
\textbf{Stability diagrams} in the $(\beta_{\mathrm{a}}, \rhoe)$ plane, with 
(a) $l_{11}=0,l_{12}=0,\gamma_{\mathrm{a}}=0$;
(b) $l_{11}=2,l_{12}=2,\gamma_{\mathrm{a}}=0$, and in the 
$(\gamma_{\mathrm{a}}, \rhoe)$ plane, with 
(c) $l_{11}=0,l_{12}=0,\beta_{\mathrm{a}}=0$;
(d) $l_{11}=2,l_{12}=2,\beta_{\mathrm{a}}=0$.
The color code and other parameter values are as in Fig.~\ref{fig:bif}.}
\label{fig:bif_rho}
\end{figure}

At fixed values of the active parameters $\beta_{\mathrm{a}}$ and
$\gamma_{\mathrm{a}},$ Fig.~\ref{fig:bif:l}a indicates that $l_{12}$ 
also suppresses the instability. However, this is not general: 
reentrant behaviour as a function of 
$l_{12}$ is possible for other parameter values, see Fig.~\ref{fig:bif:l}b.  
We briefly examined the cases including additional terms allowed by symmetry, 
such as an active transport term 
$\alpha_{\mathrm{a}} p \partial_x p$ in Eq.~\eqref{eq:const:polar:p}
or a lower-order gradient term 
$\frac{\nu_2}{2}\left(\partial_{x}p\right)^{2}$ in Eq.~\eqref{def:polar:f}
(see Fig. \ref{fig:bif:l}cd). We present in 
Figs.~\ref{fig:bif:l}cd  the stability diagrams in the $(l_{11}, l_{12})$ plane
when either $\nu_2$ or $\alpha_{\mathrm{a}}$ is non-zero.
As expected, the behavior is qualitatively the same as in Fig.~\ref{fig:bif:l}a.
Note however that $\nu_2$ suppresses somewhat the instability 
(Fig.~\ref{fig:bif:l}c).

The validity of linear stability analysis was confirmed by  
numerical simulations. As an example, we present
in Fig.~\ref{fig:bif:num}  a numerical resolution of 
Eqs.~(\ref{eq:cons:cell}-\ref{eq:cons:momentum},
\ref{eq:const:polar:kd}-\ref{eq:const:polar:p}),
supplemented with Eqs.~({\ref{def:polar:f}-\ref{def:polar:psip}}),
where we added to $\psi_{\rho}$ the fourth-order term 
 $\frac{1}{4K_4}\left(\frac{\rho-\rhoe}{\rhoe}\right)^{4}$ with $K_4=1/400$
in order  to saturate the instability.
Starting the simulation with parameters for which the uniform state 
is linearly stable, we induce the formation of a traveling wave by 
changing the value of $l_{12}$, in agreement with linear stability analysis,
see Fig.~\ref{fig:bif:l}a.

Finally, we examined the cell density dependence of the stability threshold.
We give the stability diagrams in Fig.~\ref{fig:bif_rho}. They indicate 
that a higher cell density is favorable for the instability in our model. 
In agreement with the general tendency found in Sec.~\ref{sec:app:stab:alpha0},
the instability occurs for smaller $\beta_{\mathrm{a}}$ when $w(1 - l_{12}) >0$
and for larger $\beta_{\mathrm{a}}$ when $w(1 - l_{12}) <0$.

\section{Conclusion}
\label{sec:disc}

To conclude, linear nonequilibrium thermodynamics specifies the rate of 
change of the cell density $\rate$ as a linear combination of
chemical potential, velocity divergence, activity, as well
as polarity divergence when appropriate. 
In particular, the new cross-coefficient $l_{12}$ that couples $\rate$ 
to the velocity divergence modifies cell monolayer mechanics, influencing
the velocity of advancing fronts and altering pattern-forming instabilities. 
The decomposition \eqref{eq:const:scalar:kd} agrees qualitatively 
with a large body of experiments.
Our results call for a careful quantitative comparison with experimental data, 
which will necessitate the simultaneous measurement in space
and time and at tissue scale of several fields: the cell density, the velocity, the myosin distribution, and if possible the polarity.

In the case of elastic solids, growth has been studied 
with a careful treatment of thermodynamics, up to the regime of large
deformations \cite{Ambrosi2011}.
As an advantage, our approach is easily generalizable to more complex 
rheologies including, \emph{e.g.}, orientational order parameters. 
Another advantage is that all the possible couplings are determined 
from symmetry without any ambiguity, at least in the regime of 
linear nonequilibrium.
Extensions to 2 and 3 spatial dimensions are straightforward, where 
similar constitutive equations would apply to the isotropic parts of 
the relevant tensor fields, while, for instance, the couplings between 
mechanical fields and the orientation of cell division \cite{Castanon2011}
would pertain to their deviators. Our approach is applicable \emph{in vivo},
where epithelial tissues such as the \emph{Drosophila} pupal notum and 
wings are compressible in the plane \cite{Bosveld2012,Guirao2015}.

\acknowledgments
We are pleased to acknowledge useful discussions with Shuji Ishihara, 
Jonas Ranft and Pierre Recho.  
S.Y. was supported by Grant-in-Aid for Young  Scientists
(B) (15K17737), Grants-in-Aid for Japan Society for Promotion of Science 
(JSPS) Fellows (Grants Nos. 263111), and the JSPS Core-to-Core Program 
"Non-equilibrium dynamics of soft matter and information".

%\bibliography{proliferation}
%

\newpage
\appendix

\section{Choice of fluxes and forces}
\label{sec:app:scalar}

In linear nonequilibrium thermodynamics, the choice of force \emph{vs.}
flux is arbitrary, and can be modified at will thanks to a change of 
basis by standard linear algebra (see an example below). 
The choice made here:
\begin{equation*}
\begin{array}{ccc}
\mathrm{Flux} & \leftrightarrow & \mathrm{Force}\\
\rate & \leftrightarrow & -\rho\mu\\
\left(\sigma+\pi\right) & \leftrightarrow & \partial_{x}v\\
r & \leftrightarrow & \Delta\mu
\end{array}
\end{equation*}
is one of convenience,
in order to express a poorly known quantity $\kappa$ as a function
of quantities that are either measurable ($\partial_x v$) or computable
once the free energy is given ($- \rho \mu$).
We followed standard practice concerning the other flux-force pairs,
with fluxes defined as $\sigma + \pi$ and $r$, see \emph{e.g.}
\cite{Kruse2005,Marchetti2013}.
Another approach \cite{Tlili2015} posits $\rate$ as the sum of three rates,
of cell growth, cell division, and cell death respectively. Since each 
process may be regulated by cell density 
and/or velocity divergence, and requires ATP hydrolysis for its completion, 
we expect the three rates to be functions of $\rho$, $\partial_x v$ and 
$\Delta \mu$. Only their sum $\rate$ can be specified unambiguously
by thermodynamics, Eq.~\eqref{eq:const:scalar:kd}.  

Since the choice of fluxes and forces is arbitrary at linear order, 
it is for instance possible to rewrite our constitutive equation 
(see Eq.~\eqref{eq:const:stress}) so that  
the cell density variation rate is expressed in terms 
of the stress, pressure and velocity divergence as 
\begin{equation}
\kappa-\kappa_{\mathrm{a}}
=-\frac{l_{11}}{l_{12}}\left(\sigma+\pi-\sigma_{\mathrm{a}}\right)
+\left(l_{12} + \eta \frac{l_{11}}{l_{12}} \right)\partial_{x}v.
\end{equation} 
including also the active variables $\kappa_{\mathrm{a}}$ and $\sigma_{\mathrm{a}}$.

However, such transformations into another set of forces and fluxes become
practically complicated when the Onsager coefficients depend on the 
hydrodynamic variables. 
Here, the only nonconstant Onsager coefficients that we include are related to 
activity/contractility (see the active terms in 
Eqs.~(\ref{eq:const:polar:kd}-\ref{eq:const:polar:stress})): 
the choice of $\Delta \mu$ as a force is non-trivial, but
standard in the context of active gel models \cite{Kruse2005,Marchetti2013}.

Finally, it would be possible to choose $- \mu$ as a force instead of 
$- \rho\mu$. Then the corresponding flux becomes $\kappa \rho$, and the
hydrodynamic equations would be slightly modified. We may then rewrite 
Eqs.~(\ref{eq:const:scalar:kd}-\ref{eq:const:scalar:stress}) as
\begin{eqnarray}
\label{eq:const:scalar:kd:mu}
\rate \rho&=& l_{11} \, \left(-\mu\right)+ l_{12}  \partial_{x}v+\ract \\
  \label{eq:const:scalar:stress:mu}
\sigma+\pi &=& - l_{12} \, \left(-\mu\right) 
+ \eta \, \partial_{x}v +\sact \,,
\end{eqnarray}
taking $\mu$ as a force. 
Since Onsager coefficients can have an arbitrary dependence on the 
hydrodynamic variable as far as positivity of the entropy production 
is guaranteed, if $l_{11}$, $l_{12}$ and $\ract$ are functions of the cell 
density, we may obtain
\begin{eqnarray}
\label{eq:const:scalar:kd:mu:rho}
\rate \rho&=& l_{11}' {\rho}^2\, \left(-\mu\right)+ l_{12}' \rho\, \partial_{x}v+\ract' \rho\\
  \label{eq:const:scalar:stress:mu:rho}
\sigma+\pi &=& - l_{12}' \rho\, \left(-\mu\right) + \eta \, \partial_{x}v +\sact \,.
\end{eqnarray}
A different choice of force-flux pair 
may thus lead to the same hydrodynamic equations at the price
of introducing density-dependent Onsager coefficients.

\section{Constitutive equations for a proliferating, compressible, active, and polar material}
\label{sec:app:polar}

We consider in this section a system of constant size $L$ with
periodic boundary conditions. Since the free energy of a polar material 
also depends on the polarity field and its spatial derivatives, 
$f = f(\rho,p,\partial_{x}p,\partial_{x}^{2}p )$,
the calculation of the pressure and conjugate fields must be adapted. 
This is perhaps most easily seen by considering the free energy functional
\begin{equation}
\nonumber
F=\int_{0}^{L} \mathrm{d}x \, f(\rho,p,\partial_{x}p,\partial_{x}^{2}p )\,,
\end{equation}
and computing its rate of variation:
\begin{eqnarray*}
\dot{F} & = & \frac{d}{dt} \int_{0}^{L} \mathrm{d}x \, f = 
\int_{0}^{L} \mathrm{d}x \, \partial_t f \\
&=& \int_{0}^{L} \mathrm{d}x \, \left( \DT{f}{t} - v \partial_x f \right) \\
&=& \int_{0}^{L} \mathrm{d}x \, \left( \DT{f}{t} + f \partial_x v \right) \\
 & = & \int_{0}^{L} \mathrm{d}x \
\left( f \partial_x v +
\DP{f}{\rho} \DT{\rho}{t} +
\DP{f}{p} \DT{p}{t} + \right. \\
&& \left. \DP{f}{(\partial_x p)} \DT{}{t}\left(\partial_x p\right)  
  + \DP{f}{(\partial^2_xp)} \DT{}{t}\left(\partial^2_x p\right)  \right) 
\end{eqnarray*}
% one column version
% \begin{eqnarray*}
% \dot{F} & = & \frac{d}{dt} \int_{0}^{L} \mathrm{d}x \, f = 
% \int_{0}^{L} \mathrm{d}x \, \partial_t f \\
% &=& \int_{0}^{L} \mathrm{d}x \, \left( \DT{f}{t} - v \partial_x f \right) 
% = \int_{0}^{L} \mathrm{d}x \, \left( \DT{f}{t} + f \partial_x v \right) \\
%  & = & \int_{0}^{L} \mathrm{d}x \
% \left( f \partial_x v +
% \DP{f}{\rho} \DT{\rho}{t} +
% \DP{f}{p} \DT{p}{t} + 
%  \DP{f}{(\partial_x p)} \DT{}{t}\left(\partial_x p\right)  
%   + \DP{f}{(\partial^2_xp)} \DT{}{t}\left(\partial^2_x p\right)  \right) \,.
% \end{eqnarray*}
Since
\begin{align}
\DT{}{t}\left(\partial_x p\right) & = \partial_x \DT{p}{t} 
- (\partial_x p) (\partial_x v)  \nonumber\\
\DT{}{t}\left(\partial^2_x p\right) & = \partial^2_x \DT{p}{t} 
- (\partial_x p) (\partial^2_x v) 
- 2 (\partial^2_x p) (\partial_x v) \,,
 \nonumber
\end{align}
integrations by parts yield
\begin{eqnarray*}
\dot{F} =  \int_{0}^{L} \mathrm{d}x \, 
&& \left\{   \rate \rho \DP{f}{\rho}   \right. \\ 
&& + \dot{p} \left( \frac{\partial f}{\partial p}
-\partial_{x}\left(\frac{\partial f}{\partial\left(\partial_{x}p\right)}\right)
+\partial_{x}^{2}\left(\frac{\partial f}{\partial\left(\partial_{x}^{2}p\right)}
\right) \right)    \\
 && + \partial_x v \left(
f - \rho \frac{\partial f}{\partial\rho} -\frac{\partial f}{\partial\left(\partial_{x}p\right)}\partial_{x}p 
-\frac{\partial f}{\partial\left(\partial_{x}^{2}p\right)}\partial_{x}^{2}p 
\right. \\
&&  \left. \left. \vspace*{1cm} +  \partial_{x}p \,\, \partial_x
    \left( \frac{\partial f}{\partial\left(\partial_{x}^{2}p\right)} \right)   
\right) \right\}
\end{eqnarray*}
% one column version
% \begin{eqnarray*}
% \dot{F} =  \int_{0}^{L} \mathrm{d}x \, 
% && \left\{   \rate \rho \DP{f}{\rho} + 
% \dot{p} 
% \left( \frac{\partial f}{\partial p}
% -\partial_{x} 
% \left(\frac{\partial f}{\partial\left(\partial_{x}p\right)}\right)
% +\partial_{x}^{2}
% \left(\frac{\partial f}{\partial\left(\partial_{x}^{2}p\right)}
% \right) 
% \right)  \right.  \\
%  &&  \left. + \partial_x v \left(
% f - \rho \frac{\partial f}{\partial\rho} -
% \frac{\partial f}{\partial\left(\partial_{x}p\right)}\partial_{x}p 
% -\frac{\partial f}{\partial\left(\partial_{x}^{2}p\right)}\partial_{x}^{2}p 
%  +  \partial_{x}p \,\, 
% \partial_x \left( \frac{\partial f}{\partial\left(\partial_{x}^{2}p\right)}
% \right)  \right) 
% \right\} \,.
% \end{eqnarray*}
The power of the external force on the monolayer is 
\begin{equation}
\nonumber
\Pi  =  \int_{0}^{L} \mathrm{d}x \, f_{\mathrm{ext}} v
 =  - \int_{0}^{L}  \mathrm{d}x \, v\partial_{x}\sigma 
 =  \int_{0}^{L} \mathrm{d}x \,\sigma\partial_{x}v \,.
\end{equation}
Taking into account ATP hydrolysis, the dissipation rate $R$ reads
\begin{eqnarray}
R & = & \Pi-\dot{F}+\int_{0}^{L} \mathrm{d}x \, r \Delta\mu\nonumber \\
  \label{eq:R:polar}
 & = & \int_{0}^{L}  \mathrm{d}x  \left(-\rho\mu \rate+\left(\sigma+\pi\right)\partial_{x}v+h\dot{p}+r\Delta\mu  \right).
\end{eqnarray}
with
\begin{eqnarray*}
\mu &=& \frac{\partial f}{\partial\rho} \,,\\ 
\pi & = & 
-f + \rho \frac{\partial f}{\partial\rho}
+ \left(\frac{\partial f}{\partial\left(\partial_{x}p\right)}\right)\partial_{x}p
- \partial_{x}\left(\frac{\partial f}{\partial\left(\partial_{x}^{2}p\right)}\right)\partial_{x}p 
\nonumber\\
&&+\frac{\partial f}{\partial\left(\partial_{x}^{2}p\right)}\partial_{x}^{2}p \,,\\
% one column version
% +\frac{\partial f}{\partial\left(\partial_{x}^{2}p\right)}\partial_{x}^{2}p \,,\\
h &=& -\frac{\partial f}{\partial p}
+\partial_{x}\left(\frac{\partial f}{\partial\left(\partial_{x}p\right)}\right)
-\partial_{x}^{2}\left(\frac{\partial f}{\partial\left(\partial_{x}^{2}p\right)}\right) \,.
\end{eqnarray*}
Using \eqref{def:polar:f} as the free energy density, we find
\begin{eqnarray*}
\mu&=&\psi'_{\rho}(\rho) + w \partial_x p \,,\\
\pi&=& - f + \rho \psi'_{\rho}
{+ 2 w \rho \partial_x p}
% two column version
% \nonumber\\
% &&+\nu_4 \left[  \left(\partial_{x}^{2}p\right)^{2}
% - \left(\partial_{x}p \right) \left(\partial_{x}^3p \right)\right] \,,\\
+\nu_4 \left[  \left(\partial_{x}^{2}p\right)^{2}
- \left(\partial_{x}p \right) \left(\partial_{x}^3p \right)\right] \,,\\
h&=&-\psi'_{\mathrm{p}}\left(p\right)+w \partial_{x} \rho
-\nu_4 \partial_{x}^{4}p \,.
\end{eqnarray*}

\section{Linear stability analysis}
\label{sec:app:stab}

\subsection{General case}
\label{sec:app:stab:general}

Setting for simplicity $\ract = 0$, we study the linear stability of the 
uniform state  $(\rho_0, p_{0}, v_0)$ with $\rho_0 = \rhoe$ and 
$v_0 = f_{\mathrm{a}} p_0/\xi$. A perturbation of small amplitude with 
wave number $q$ reads 
\begin{equation}
\label{eq:linear:polar:pert}
\left(\rho,p,v\right) =
\left(\rhoe,p_{0},v_0\right)
+ \left(\delta\rho,\delta p,\delta v\right)\, e^{st-iqx}
\end{equation}
with a growth rate $s(q)$. Taking into account 
Eqs.~(\ref{eq:const:polar:kd}-\ref{eq:const:polar:stress}) and 
(\ref{def:polar:f}-\ref{def:polar:psip}), we find
at linear order and with similar notations:
\begin{eqnarray*}
\delta(\rho\mu)& =  & \frac{1}{K\rhoe} \, \delta\rho - iqw\rhoe \, \delta p \\
\delta\kappa&  = &  
-\frac{l_{11}}{K\rhoe}\delta\rho + iq \, 
( w l_{11} \rhoe  - \gamma_{\mathrm{a}}) \delta p
-iq l_{12}  \delta v 
\end{eqnarray*}
\vspace*{-0.5cm}
\begin{eqnarray*}
\hspace*{0.5cm}\delta\pi& = & \frac{1}{K\rhoe} \, \delta\rho - iqw {\rhoe} \,\delta p \\
\delta\sigma&  = &  
- \frac{L_{12}}{K \rhoe} \, \delta\rho  
-iq \left( \beta_{\mathrm{a}} %+ (l_{12} - 1) 
- L_{12} w {\rhoe} \right) \, \delta p
-iq  \eta \delta v\\
\delta h& = & -iqw \, \delta\rho -\left(2a_{2}+\nu_{4}q^{4}\right) \,\delta p \,,
\end{eqnarray*}
where $L_{12}=1-l_{12}$. Substituting into 
Eqs.~(\ref{eq:cons:cell}-\ref{eq:cons:momentum}-\ref{eq:const:polar:p}), 
the amplitudes of perturbations obey at linear order:
\widetext
\begin{eqnarray}
\label{eq:linear:polar:l1}
\left(s+ \frac{l_{11}}{K} - i q v_0\right) \, \delta\rho
- iq {\rhoe}\left(l_{11}w\rhoe-\gamma_{\mathrm{a}}\right) \, \delta p
-iq\rhoe L_{12}  \, \delta v 
&=& 0 \\
\label{eq:linear:polar:l2}
\frac{iq}{K\rhoe} L_{12}  \, \delta\rho + \left( f_{\mathrm{a}} +  q^{2} 
\left( L_{12} w \rhoe- \beta_{\mathrm{a}} \right) \right) \, \delta p
- \left( \xi + \eta q^2 \right) \,\delta v
&=&0 \\
\label{eq:linear:polar:l3}
iq\Gamma_{\mathrm{p}}w\,\delta\rho
+ \left(s+\Gamma_{\mathrm{p}}\left(2a_{2} + \nu_{4}q^{4}\right) 
   -iq  v_0 \right)  \,\delta p
&=&0 
\end{eqnarray}
\twocolumngrid 
When $\mathrm{Re}(s)<0$  (respectively $\mathrm{Re}  (s)>0$), the uniform state 
$\rho(x)=\rho_{0}$, $p(x)=p_{0}$, $v(x)=v_0$, is stable (respectively unstable).
By evaluating numerically the largest real part of $s$, we obtain the 
stability diagrams plotted in figures.

\subsection{Analytical calculation in a simple case}
\label{sec:app:stab:alpha0}

Setting $\eta = 0$, $l_{11}=0$, $\gamma_{\mathrm{a}}=0$ and $f_{\mathrm{a}}=0$, an 
analytical expression of the stability threshold can be obtained. 
Contrary to the general case, the instability is here 
\emph{stationary}, but this calculation is useful to understand the 
bifurcation diagrams in Fig.~\ref{fig:bif}.
The growth rate $s$ is a solution of the polynomial equation
\begin{equation*}
\xi s^{2}+B\left(q^{2}\right)s+C\left(q^{2}\right)=0
\end{equation*}
with:
\begin{eqnarray*}
B\left(q^{2}\right)&=&\Gamma_{\mathrm{p}}\xi \nu_{4} \, q^{4}+
\frac{(L_{12})^{2}}{K}\,q^{2}+2a_{2}\Gamma_{\mathrm{p}}\xi\\% \, \geq 0\\
 C\left(q^{2}\right)
 % & =& \frac{\Gamma_{\mathrm{p}}}{K} (L_{12})^{2} \nu_4 \, q^{6}- \Gamma_{\mathrm{p}} w \rhoe L_{12}  \left(L_{12} \, \rhoe w-\beta_{\mathrm{a}}\right) \, q^{4}+ 2 a_{2} \frac{\Gamma_{\mathrm{p}}}{K}  (L_{12})^{2} \,q^{2}\nonumber \\
 & =&  \frac{\Gamma_{\mathrm{p}}}{K}  (L_{12})^{2} \nu_4 \, q^{2} \, \times \nonumber\\
&&
\left[q^{4}-
 \frac{K \rhoe}{\nu_4}  w \left( w \rhoe- \frac{\beta_{\mathrm{a}}}{L_{12}} \right) \, q^{2}
 +\frac{2a_{2}}{\nu_{4}}\right]
\end{eqnarray*}
Since $B(q^2) \geq 0$, the instability occurs when the minimum 
of $C\left(q^{2}\right)$ with respect to $q^2$ becomes negative, provided that
\begin{equation*}
w\left(w\rhoe-\frac{\beta_{\mathrm{a}}}{L_{12}}\right)>0.
\end{equation*}
Since the minimum of $C\left(q^{2}\right)$ is, up to a positive factor, 
proportional to 
\begin{equation*}
-\frac{1}{4} \left( \frac{K \rhoe w}{\nu_4} \right)^2  \left(  w \rhoe - \frac{\beta_{\mathrm{a}}}{L_{12}} \right)^2
+2\frac{a_{2}}{\nu_4} \,,
\end{equation*}
the conditions for an instability are equivalent to
\begin{equation}
\label{eq:linstab:simple:cond}
w\left(w\rhoe-\frac{\beta_{\mathrm{a}}}{L_{12}}\right)
>\frac{2}{ \rhoe K}\sqrt{2a_{2}\nu_{4}} \,.
\end{equation}
Eq.~\eqref{eq:linstab:simple:cond} allows to define a threshold value 
$\beta_{\mathrm{a}}^c$ of the active parameter $\beta_{\mathrm{a}}$:
\begin{equation*}
\beta_{\mathrm{a}}^c = L_{12}\left(w\rhoe-\frac{2}{w \rhoe K}\sqrt{2a_{2}\nu_{4}}\right) \,,
\end{equation*}
with two cases depending on the sign of the product $L_{12}w = (1-l_{12})w$.

If $(1-l_{12})w > 0$ (respectively $(1-l_{12})w < 0$ ), the instability takes 
place when $\beta_{\mathrm{a}} < \beta_{\mathrm{a}}^c$ 
(respectively $\beta_{\mathrm{a}} > \beta_{\mathrm{a}}^c$).
This result agrees with the general tendency found numerically, 
and which holds in the general case 
$l_{11}\neq0$, $f_{\mathrm{a}}\neq0$, $\gamma_{\mathrm{a}}\neq0$,  that smaller 
(respectively larger) $\beta_{\mathrm{a}}$ is favorable for the instability when 
$(1 - l_{12})w >0$ (respectively $(1 - l_{12})w <0$).

The limit case $L_{12} = 0$, $l_{12} = 1$ leads to marginal stability 
(assuming as above that  $\eta = l_{11}= \gamma_{\mathrm{a}}=f_{\mathrm{a}} = 0$), 
with the growth rates $s = 0$, 
$s = -\Gamma_{\mathrm{p}} \left( 2 a_{2} + \nu_4 q^4 \right)$, 
and may require a non-linear analysis.

\begin{figure}[!h]
\showfigures{
\includegraphics[scale=0.2]{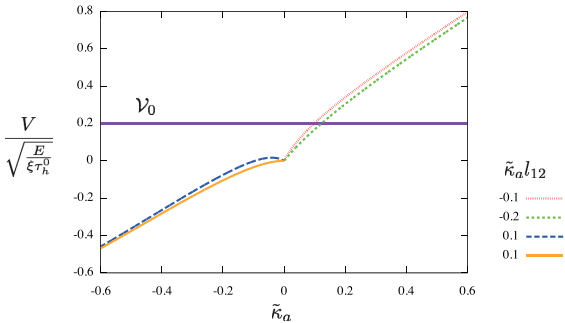}
}
\caption{\textbf{Dimensionless front velocity}
$\mathcal{V} = V/\sqrt{\frac{E}{\xi\, \tauh^0}}$ as a function of 
$\tilde{\kappa}_{\mathrm{a}}$ at fixed 
$\tilde{\kappa}_{\mathrm{a}} l_{12} = -0.2, -0.1, 0.1, 0.2$,
assuming $l_{12} \le 0$, with $\alpha_0 = 0.2$, $\etascale = 10^{-2}$.
The solid line indicates the reference velocity at $l_{12} = \ract = 0$,
$\mathcal{V}_0 = \alpha_0/\sqrt{1 + \etascale} \simeq \alpha_0  = 0.2$. 
}
\label{fig:V2}
\end{figure}

\begin{figure}[!h]
\vspace*{0.5cm}
\showfigures{
\includegraphics[scale=0.5]{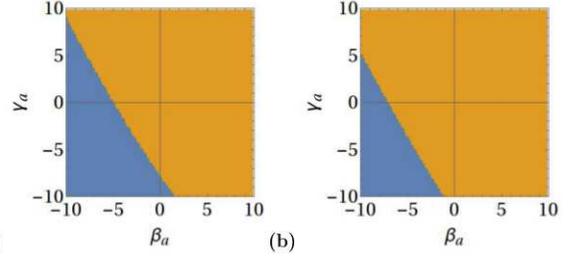}
}
\caption{
\textbf{Stability diagrams} in the plane $(\beta_{\mathrm{a}}, \gamma_{\mathrm{a}})$ 
of active control parameters for 
(a) $l_{11} = 0$, $l_{12} = -2$; 
(b) $l_{11} = 2$, $l_{12} = -2$. 
The color code and other parameter values are as in Fig.~\ref{fig:bif:l}.
}
\label{fig:bif:2}
\end{figure}

\begin{figure}[!h]
\showfigures{
\includegraphics[scale=0.5]{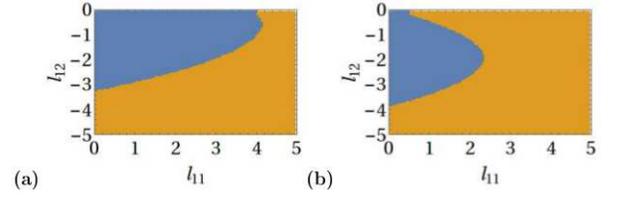}
}
\caption{
\textbf{Stability diagrams} in the plane $(l_{11}, l_{12})$, with 
(a) $\beta_{\mathrm{a}}=-5$, $\gamma_{\mathrm{a}}= -5$;
(b) $\beta_{\mathrm{a}}=-10$, $\gamma_{\mathrm{a}}=5$.
The color code and other parameter values are as in Fig.~\ref{fig:bif:l}.
}
\label{fig:bif:3}
\end{figure}

\section{Cases with $l_{12}$ negative}
\label{sec:app:negativel12}

As mentioned in the main text, experiments suggest that the cross-coupling 
$l_{12}$ is positive. Since linear nonequilibrium thermodynamics cannot 
exclude a negative sign for $l_{12}$, we briefly examine in this section, 
for each example, cases with a negative $l_{12}$

\subsection{Front velocity for $l_{12} < 0$}
\label{sec:app:negativel12:frontvelocity}

Fig.~\ref{fig:V2} shows how the dimensionless front velocity $\mathcal{V}$
depends on the active rate $\tilde{\kappa}_{\mathrm{a}}$ at fixed, small 
$\ract l_{12}$, when $l_{12} < 0$ and $\alpha_0 = 0.2$. $\mathcal{V}$ is a 
monotonically increasing function of $\tilde{\kappa}_{\mathrm{a}}$ except  for
$\tilde{\kappa}_{\mathrm{a}}$ close to $0$ and negative. A difference 
with the case $l_{12}\ge0$ examined in the main text is the absence of a 
sharp peak, observed near $l_{12}=1$ in Fig.~\ref{fig:V}. 
Note that $\mathcal{V}$ rapidly changes sign to become negative fo 
$\tilde{\kappa}_{\mathrm{a}} < 0$, $l_{12} < 0$.

\subsection{Stability analysis for $l_{12} < 0$}
\label{sec:app:negativel12:stab}

We give the stability diagrams in the plane $(\beta_{\mathrm{a}},\gamma_{\mathrm{a}})$ 
for $l_{11}=0$, $l_{12}=-2$ (Fig.~\ref{fig:bif:2}a) and $l_{11}=2$, $l_{12}=-2$ 
(Fig.~\ref{fig:bif:2}b). A smaller $\beta_a$ is favorable for the instability 
in both cases with $l_{12}<0$, in accord with the general tendency for 
$w (1-l_{12})>0$ explained above. By comparing  Fig.~\ref{fig:bif:2}a and 
Fig.~\ref{fig:bif:2}b, we see that the instability is suppressed due to 
$l_{11}$, as has also been observed in the main text 
in several cases with $l_{12}\geq 0$. 
Finally, we also observe reentrant behavior 
as a function of $l_{12}$ in the region $l_{12}<0$, see Fig.~\ref{fig:bif:3}.

\end{document}